\PassOptionsToPackage{dvipsnames}{xcolor}

\documentclass[sigconf]{acmart}
\AtBeginDocument{%
  }

\setcopyright{acmlicensed}
\copyrightyear{2025}
\acmYear{2025}
\acmDOI{XXXXXXX.XXXXXXX}
\acmConference[EASE 2025]{The 29th International Conference on Evaluation and Assessment in Software Engineering}{17–20 June, 2025}{Istanbul, Türkiye}

\usepackage{amsmath,amssymb,amsfonts}
\usepackage{graphicx}
\usepackage{textcomp}
\usepackage{marvosym}
\usepackage{booktabs}   %about table
\usepackage{multirow} 
\usepackage{colortbl}
\usepackage[bottom,flushmargin]{footmisc}
\usepackage{breakurl}

\usepackage{bbding}
\usepackage{threeparttable}

\usepackage[ruled,vlined,linesnumbered]{algorithm2e}
\usepackage{fancyhdr}

\hypersetup{
    colorlinks=true,
    linkcolor=black,
    citecolor=black,
    urlcolor=blue,
}
\usepackage{listings}
\definecolor{code-red}{RGB}{255 176 187}
\definecolor{code-green}{RGB}{179 247 179}
\definecolor{light-gray}{RGB}{211 211 211}

\begin{document}

% \title{How Well Do Large Language Models Serve as End-to-End Secure Code Producers?}
\title{How Well Do Large Language Models Serve as End-to-End Secure Code Agents for Python?}

%二级标题是不是改成全部大写(√)
%在prompt部分介绍为什么只用一种(√)
%正文中的producer要改成agent相关(√)
%fig 11 改成椭圆(√)
%fig 13看看是不是添加更多的信息
%其余按照列出的要点更改
    %professor那个表述改一改(√)
    %比较表格改成比language；scenarios要不要加个注释 (√)
    %RQ1按原定移动大部分到附录之中(√)
    %修改原来的流程图（review的对比关系，图的高度的协调性）；(√)
    %把算法图换成伪代码；(√)
    %fig 12更加突出agent的特性（加个小人？），是不是改成teaser(√)
    %摘要精简(1)，写上Python(√)

% M. F. Rabbi, A. I. Champa, M. F. Zibran, and M. R. Islam, “AI writes we analyze: The ChatGPT python code saga,” in Proceedings of ACM International Conference on Mining Software Repositories (MSR), 2024

\author{Jianian Gong}
\email{jianian_gong@buaa.edu.cn}
\affiliation{%
  \institution{School of Computer Science and Engineering, Beihang University}
  \city{Beijing}
  % \state{Beijing}
  \country{China}
}

\author{Nachuan Duan}
\email{nc_duan@buaa.edu.cn}
\affiliation{%
  \institution{School of Computer Science and Engineering, Beihang University}
  \city{Beijing}
  % \state{Beijing}
  \country{China}
  }

\author{Ziheng Tao}
\email{tzh588577@163.com}
\affiliation{%
  \institution{School of Computer Science and Engineering, Beihang University}
  \city{Beijing}
  % \state{Beijing}
  \country{China}
  }

\author{Zhaohui Gong}
\email{gzh1198073707@163.com}
\affiliation{%
  \institution{School of Computer Science and Engineering, Beihang University}
  \city{Beijing}
  % \state{Beijing}
  \country{China}
  }

\author{Yuan Yuan}
\authornote{Corresponding author.}
\email{yuan21@buaa.edu.cn}  
\affiliation{%
  \institution{School of Computer Science and Engineering, Beihang University}
  \institution{State Key Laboratory of Software Development Environment}
  \institution{Zhongguancun Laboratory}
  \city{Beijing, 100191}
  % \state{Beijing}
  \country{China}
  }

\author{Minlie Huang}
\email{aihuang@tsinghua.edu.cn}  
\affiliation{%
  \institution{Tsinghua University}
  \institution{Zhongguancun Laboratory}
  \city{Beijing}
  % \state{Beijing}
  \country{China}
  }


\begin{abstract}
The rapid advancement of large language models (LLMs) such as GPT-4 has revolutionized the landscape of software engineering, positioning these models at the core of modern development practices. To fully realize their potential in producing secure source code autonomously, LLMs must not only generate code but also identify and repair vulnerabilities in their outputs, thereby improving security iteratively. Despite growing prominence, LLMs' effectiveness in performing such \textbf{end-to-end tasks} remains unexplored. This paper bridges this gap by systematically investigating the capability of LLMs to generate source code, evaluate \textbf{their own outputs} for vulnerabilities, and apply necessary repairs to improve the security of \textbf{their self-generated code}.

Specifically, we studied the ability of GPT-3.5 and GPT-4 to identify and repair vulnerabilities in the code generated by four popular LLMs including themselves\ (GPT-3.5, GPT-4, Code Llama, and CodeGeeX2). By manually or automatically reviewing 4,900 pieces of code, our study reveals that: (1)\ LLMs generate over 75\% vulnerable Python code in given scenarios; (2)\ LLMs such as GPT-3.5 and GPT-4 are unable to precisely identify vulnerabilities in the code they generated; (3)\ GPT-3.5 and GPT-4 can achieve 33.2\%$\sim$59.6\% success rates in repairing the insecure code produced by the 4 LLMs, but they both perform poorly when repairing self-produced code, indicating self-repair ``blind spots". To address the limitation of a single round of repair, we developed a lightweight tool using LLMs as agents to construct safer source code through an iterative repair procedure based on the insights gained from our study. Experiments show that, assisted by semantic analysis engines, our tool significantly improves the success rates of repair to 65.9\%$\sim$85.5\%.
\end{abstract}




\maketitle

\begin{table*}[t]    
\renewcommand{\arraystretch}{1.5}
\caption{Comparison with representative related work evaluating the security of LLM-generated code}
\centering
\begin{threeparttable}
\footnotesize
\begin{tabular}{c|c|c|c|c|c}
\hline
Related Work\tnote{1}&Studied Language(s)&Automated Review&Manual Review&\#Studied CWEs&\#Studied LLM(s)\\
\hline\hline
Pearce et al.\cite{pearce2022asleep}&Python, C, Verilog&CodeQL&\Checkmark&25&1\\
\hline
Khoury et al.\cite{khoury2023secure}&C, C++, Python, HTML, Java&\XSolidBrush&\Checkmark&14&1\\
\hline
Asare et al.\cite{asare2023github}&C, C++&\XSolidBrush&\Checkmark&6&1\\
\hline

Sandova et al.\cite{sandoval2023lost}&C&\XSolidBrush&\Checkmark&6 &1\\
\hline

Fu et al.\cite{fu2023security}&Python, JavaScript&CodeQL, Bandit, ESLint&\XSolidBrush&16&3\\

\hline

% Hajipour et al.\cite{hajipour2024codelmsec}&Python&CodeQL&\XSolidBrush&45 CWEs&5\\
%\hline cite: 24

% Siddiq et al.\cite{siddiq2024sallm}&Python&CodeQL&\XSolidBrush&45 CWEs&5\\

Liu et al.\cite{liu2024no}&C, C++, Java&CodeQL&\XSolidBrush&17&1\\
\cellcolor{light-gray}Our work\cellcolor{light-gray}&\cellcolor{light-gray}Python&\cellcolor{light-gray}CodeQL, Bandit&\cellcolor{light-gray}\Checkmark&\cellcolor{light-gray}\textbf{69}&\cellcolor{light-gray}4\\
\hline
\end{tabular}

 \begin{tablenotes}
        \footnotesize
        \item[1] Only studies which are based on code generated by LLMs and classify vulnerabilities according to CWEs are included in this comparison

      \end{tablenotes}
  \end{threeparttable}
  \label{tab:comparison}
\end{table*}

% \begin{table}[t]    
% \renewcommand{\arraystretch}{1}
% \caption{Comparison between our work and comparable related work in terms of RQ1$\sim$RQ3}
% \centering
% \begin{threeparttable}
% \footnotesize
% \begin{tabular}{|c|c|c|}
% \hline
% Related Work\tnote{1}&Studied Language(s)&Studied CWEs\\
% \hline\hline
% Pearce et al.\cite{pearce2022asleep}&Python, C, and Verilog&25 CWEs\\
% \hline
% Khoury et al.\cite{khoury2023secure}&C, C++, Python, HTML and Java&14 CWEs\\
% \hline
% Asare et al.\cite{asare2023github}&C, C++&6 CWEs\\
% \hline
% Sandova et al.\cite{sandoval2023lost}&C&6 CWEs\\
% \hline
% Toth et al.\cite{toth2024llms}&PHP&3 CWEs\\
% \hline
% % Tihanyi et al.\cite{tihanyi2024neutral}&C&10 CWEs\\
% \hline
% \cellcolor{light-gray}Our work\cellcolor{light-gray}&\cellcolor{light-gray}\textbf{Python}&\cellcolor{light-gray}\textbf{69 CWEs}\\
% \hline
% \end{tabular}

%  \begin{tablenotes}
%         \footnotesize
%         \item[1] Only studies whose generated code snippets have all gone through manual review to achieve ground-truth are compared here
%       \end{tablenotes}
%   \end{threeparttable}
%   \label{tab:comparison}
% \end{table}

\section{INTRODUCTION}
Transformer-based large language models have fundamentally reshaped software engineering in recent years. As LLM-powered programming assistants such as GitHub Copilot widely adopted by IT companies and individual developers, traditional developer-centered software engineering (i.e., Software 1.0 and Software 2.0 where all source code is manually written) is rapidly evolving to LLM-centered software engineering (i.e., Software 3.0 where most source code is generated by AI). According to research by GitHub, its Copilot contributes to more than 46\% of all source code that its 1.3 million users have developed \cite{economicimpact}.  

As more LLM-generated code is accepted by developers and thus becomes part of the software, its security is gaining increasing concern. Previous studies \cite{pearce2022asleep, khoury2023secure, nair2023generating, asare2023github, sandoval2023lost, fu2023security, liu2024no} have revealed that although large language models are capable of generating functionally correct code, they are not free of security vulnerabilities. Therefore, raw LLM-generated source code cannot be trusted to be deployed in security-sensitive scenarios. 

The increasingly significant role of LLMs in software engineering, coupled with disturbing vulnerabilities in the code they generate, compels us to explore methods for producing safer code with LLMs. Standard end-to-end software development practices, which include writing, reviewing, and refactoring code, result in significantly higher quality compared to focusing solely on code writing \cite{refactorimprove}. 
This inspires us to consider: \textit{How well do large language models function as comprehensive, end-to-end secure code agents?} In the context of this research, \textit{end-to-end} refers to a process wherein a code snippet is not only generated but also undergoes review and, if necessary, remediation of identified vulnerabilities before being presented to users. To achieve this level of competence, LLMs must demonstrate proficiency in three key areas: generating functionally correct code, conducting thorough self-reviews of their output, and effectively repairing any detected security flaws.

Our investigation focuses specifically on evaluating these capabilities within the \textbf{Python} programming ecosystem. This choice is motivated by two primary factors: (1) Python's widespread adoption in web development, a domain where security considerations are paramount; and (2) the scarcity of existing research focusing on the security implications of AI-generated Python code.

Hence, we seek answers to the following research questions:

\textbf{RQ1: How do LLMs perform when generating Python code in security-sensitive scenarios?} This research question aims to address the lack of exclusive research on the security of LLM-generated Python code. Additionally, the identified insecure code in this RQ will be used in subsequent research questions.

\textbf{RQ2: How effective are LLMs in identifying LLM-generated code vulnerabilities?} This research question seeks to investigate LLMs' potential to self-review their generated code, as accurate identification of vulnerabilities is crucial for mitigating weaknesses in the source code.

\textbf{RQ3: How effective are LLMs in repairing LLM-generated code vulnerabilities?} This RQ investigates LLMs' capability to self-repair their generated code, which is the final and most crucial step in constructing secure source code.

\textbf{RQ4: How effective is an iterative strategy in improving LLMs' repair capability?} A single round of vulnerability detection and repair may not be sufficient to eliminate weaknesses in code. To address this, we introduce an iterative strategy that allows LLMs to repair the generated code through multiple attempts with feedback information provided. \textbf{This research question synthesizes RQ1 through RQ3, serving as the focal point of our study.}

Previous studies such as \cite{pearce2022asleep, zhou2024large, pearce2023examining} have examined, respectively, the ability of LLMs in generating secure code, detecting vulnerabilities in real-world code, or repairing them, similar to our effort in RQ1, RQ2, or RQ3. However, they suffer from limitations such as a narrow range of CWEs and restricted code review methods. In terms of RQ1, a comparison with representative work is detailed in Table \ref{tab:comparison}. \textbf{Furthermore, there has been no comprehensive pipeline study integrating the processes of generation, detection, and repair to evaluate the potential of LLMs to produce secure code in an end-to-end manner, a gap that our RQ4 aims to address.}

Our novel contributions are as follows:

(1) We present a systematic study evaluating the performance of 4 LLMs in generating secure Python code across 67 CWEs, offering a more extensive analysis than previous research as well as addressing the lack of relevant research focusing on Python. Overall, we found that the 4 tested LLMs generated over 75\% vulnerable code on the SecurityEval benchmark. Meanwhile, we made several novel discoveries such as self-repair "blind spots" of LLMs.

(2) To the best of our knowledge, we are the first to conduct a study on LLMs' efficacy in judging and fixing \textbf{insecure code generated by LLMs themselves}, while previous work focuses primarily on manually written code in open-source projects. Our experiments revealed that GPT-3.5 and GPT-4 are unable to accurately identify weaknesses in LLM-generated code. While they achieve certain success rates in repairing the insecure code produced by the 4 LLMs (33.2\% and 59.6\% respectively), they perform poorer fixing self-produced code.

(3) We designed a tool using LLMs as agents to produce much safer code through a simple feedback-driven iterative self-repair process. The tool achieves 65.9\%$\sim$85.5\% success rates of repair when using GPT-3.5 or GPT-4 as the code generator and repairer. 

For the purpose of open science, we released the artifacts of this paper in our repository\footnote{\href{https://github.com/jianian0318/LLMSecureCode}{https://github.com/jianian0318/LLMSecureCode}}.

\section{STUDY SETTING}

\subsection{Studied Dataset}
\begin{figure}[t]
\includegraphics[width=0.8\linewidth, height=1.35cm]{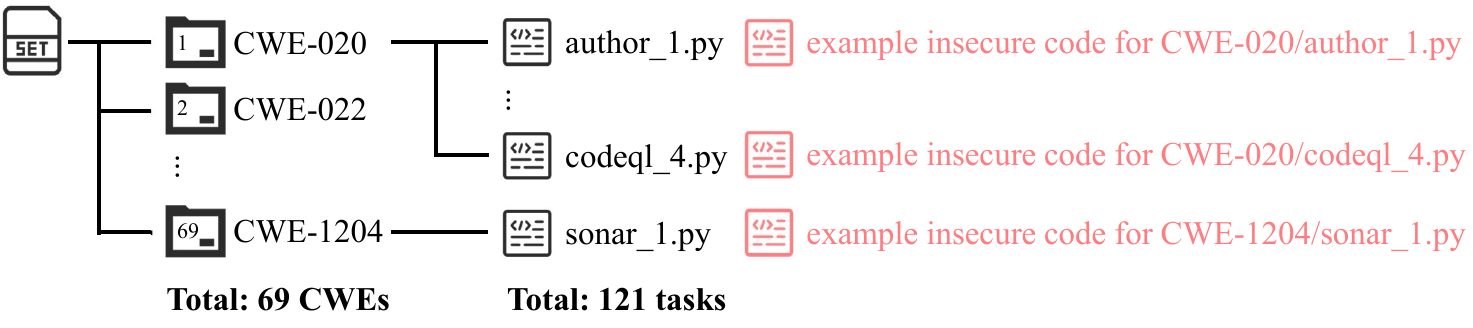}
    \caption{The structure of SecurityEval}
    \label{fig:dataset}
\end{figure}

SecurityEval is a dataset dedicated to measuring the security of machine learning-based code generation techniques\cite{siddiq2022securityeval}. In its current version, the dataset contains 121 code completion tasks in Python, all of which are exposed to the potential risk of a certain CWE. Overall, these tasks cover 69 CWEs. The dataset also contains an example vulnerable solution for each generation task. Fig. \ref{fig:dataset} depicts the structure of SecurityEval.

We chose SecurityEval because (1) it was specially designed for evaluating the security of LLM-generated code, which is highly relevant to this work; (2) it covers the widest (by the time we conclude this research) range of 69 CWEs (121 scenarios), making our work more comprehensive than previous relevant ones \cite{pearce2022asleep, khoury2023secure, nair2023generating}; (3) all of its code generation tasks are presented in Python, which matches our research goal.

\subsection{Studied Large Language Models}

% Large language models capable of code processing tasks can be divided into 2 categories: general language models represented by the GPT family and specialized models that are specifically pre-trained on code (i.e., code language models)  \cite{zhang2023unifying}.  For better representation, we choose 2 popular models for our work from each category. 

\textbf{(1) GPT-3.5} is a well-known large language model with 175B parameters proposed by OpenAI in 2022. It is capable of understanding and generating natural language text or code. We use gpt-3.5-turbo-0125 in the study. We choose GPT-3.5 because it is one of OpenAI's most popular models which powered ChatGPT to gain more than 100 million active users in 2023 \cite{chatgptgrow}.

\textbf{(2) GPT-4} is a multi-modal large language model introduced by OpenAI in 2023, capable of producing more accurate and secure responses \cite{openai2024gpt4}. The specific model used in the study is gpt-4-0613. GPT-4 is included in this work because it is one of OpenAI's most advanced models, showing remarkable performance on code generation benchmarks such as HumanEval \cite{humaneval}.

\textbf{(3) Code Llama} is a code language model introduced by Meta in 2023. It was specially trained for code-relevant tasks based on the LLaMA 2 model, with excellent performance in code generation \cite{rozière2024code}. The specific model used in the study is Code Llama-70B. We include Code Llama in our research because it is an outstanding representation of open-source code language models.

\textbf{(4) CodeGeeX2} is an open-source code language model with 6B parameters, developed jointly by Tsinghua University and Zhipu AI. It is based on the ChatGLM2 architecture and tailored for code-relevant tasks such as code generation, completion, interpretation, and cross-programming language translation \cite{CodeGeeXCopilot}\cite{codegeexintro}. CodeGeeX2 was selected for our study because of its strong performance among lightweight LLMs for code \cite{zheng2023codegeex}.  

Our research exclusively considers GPT-based large language models due to their state-of-the-art performance and their widespread use as programming assistants such as GitHub Copilot. 

\subsection{Methodology}

\subsubsection{\textbf{Methodology for RQ1$\sim$RQ3}}
\begin{figure}[t]
    \includegraphics[width=\linewidth, height=8.6cm]{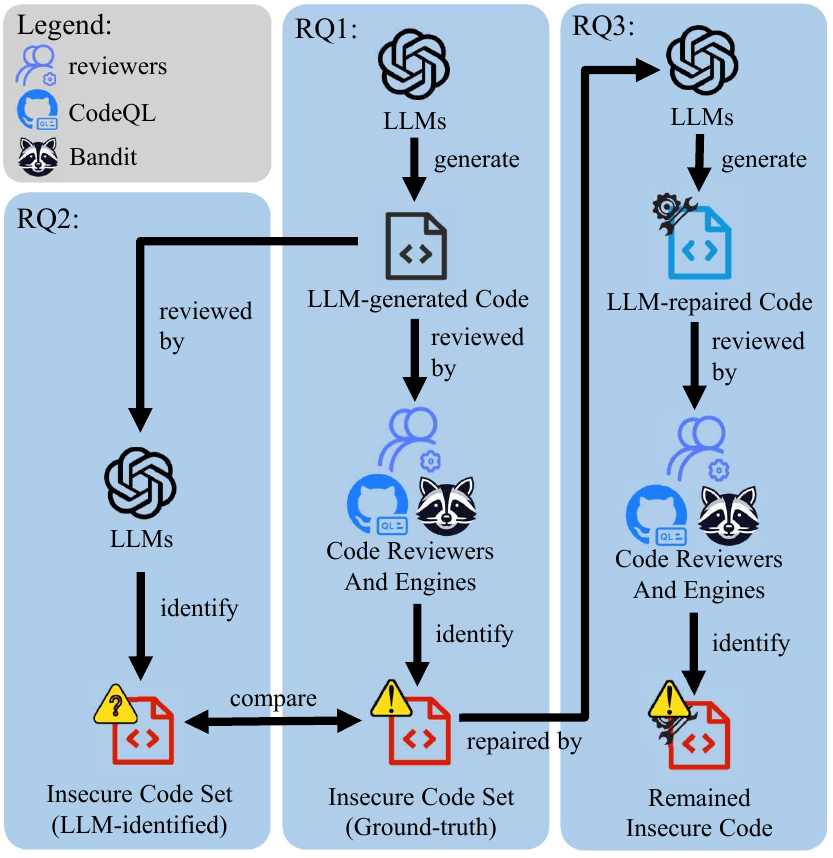}
    \caption{Workflow of RQ1$\sim$RQ3}
    \label{fig:workflow}
\end{figure}

Fig. \ref{fig:workflow} shows the overall workflow of RQ1$\sim$RQ3, which goes in a pipeline fashion. The overall procedure aims to assess LLMs' end-to-end capability to generate secure code.

In RQ1, we prompt GPT-3.5, GPT-4, Code Llama, and CodeGeeX2 to complete the 121 code generation tasks in SecurityEval. Results with syntax errors will be re-generated until free of errors. We then review them to determine whether they are with the specific CWE vulnerabilities (explained in \ref{review}) and draw conclusions. 

In RQ2, we prompt large language models to inspect every piece of code generated in RQ1 for the presence of the corresponding CWE vulnerability. We then use the review results from RQ1 as the ground truth to evaluate the LLMs' ability to identify weaknesses in their self-produced code.

In RQ3, the vulnerable code identified in RQ1 will be provided to LLMs for repair, along with information about its corresponding CWE. The Repaired code produced by these models will then undergo the review procedure again. We will assess the LLMs' capability of repairing self-produced code based on these review results.

\subsubsection{\textbf{Algorithm for RQ4}}\label{algorithm}
To exploit the potential of LLMs in producing safer code, we design an iterative repair algorithm for RQ4, as shown in Algorithm \ref{algorithm}. Taking code generation tasks ($T$) as input, it initially utilizes a LLM to generate code (\verb|LLM_gen()|) according to given task descriptions. Subsequently, it employs an automatic tool (\verb|scanner()|, CodeQL and Bandit are used in RQ4) to scan the generated code for vulnerabilities. If a piece of code is free of weaknesses, it is output as secure code. Otherwise, it is returned to the LLM for repair (\verb|LLM_repair()|), with CWE information provided ($INFO_{CWE}$). This scan-and-repair process is conducted iteratively until the code is secure or the predefined maximum number of iterations ($k$) is reached.

It is worth noting that the ``scanner" in the algorithm is not predefined. Initially, we planned to use LLMs for vulnerability self-detection. However, as discovered in RQ2, LLMs are unable to provide reliable detection results. Alternatively, semantic analysis engines such as CodeQL and Bandit have been adopted.

% \begin{figure}[t]
% % \setlength{\abovecaptionskip}{-0.2cm}
%     % \setlength{\belowcaptionskip}{-0.5cm}
%     \includegraphics[width=\linewidth]{paper-figure/newAlgorithm.pdf}
%     \caption{Algorithm for RQ4}
%     \label{fig:algorithm}
% \end{figure}

\begin{algorithm}
\SetAlgoLined
\caption{The algorithm for secure code agent}
\label{algorithm}
\SetKwFunction{CodeQL}{CodeQL}
\SetKwFunction{Manual}{Manual}
\SetKwFunction{Bandit}{Bandit}
\SetKwFunction{LLMGenerate}{LLM\_gen}
\SetKwFunction{LLMRepair}{LLM\_repair}
\SetKwFunction{Vul}{Vul}
\SetKwFunction{Scanner}{scanner}
\SetKwFunction{add}{.add}
\SetKw{And}{and}
\SetKw{Or}{or}
\SetKw{In}{in}
\SetKwComment{Comment}{{// }}{}

\KwIn{Set of code generation tasks, $T$; Max iteration, $k$}
\KwOut{Set of secure code snippets, $S$}
% \SetKwProg{Fn}{Function}{:}{end}
%     \Fn{\Vul {$p$}}{
%     \Return \Manual{$p$} \Or \CodeQL{$p$} || \Bandit{$p$}
%     }
$S \gets \emptyset$, $V \gets \emptyset$, $i \gets 1$ \\
\For{$t$ \In $T$} { 
    $p \gets$ \LLMGenerate{$t$},\ \ $(failed,\ INFO_{CWE}) \gets \Scanner{p}$ \\
    \eIf{$failed$}{$V \gets V \cup \{(p,\ INFO_{CWE})\}$}{$S \gets S \cup \{p\}$}
}
\While{$i\leq k$ \And $V \neq \emptyset$} {
    $V_{temp} \gets \emptyset$\\
    \For{$(v,\ INFO_{CWE})$ \In $V$} {
     % $V\gets V - \{(v,\ INFO_{CWE})\}$\\
     $v_{fix} \gets$ \LLMRepair{$v,\ INFO_{CWE}$}\\
     $(failed,\ INFO_{CWE}) \gets$ \Scanner{$v_{fix}$} \\
    \eIf{$failed$}{$V_{temp} \gets V_{temp} \cup \{(v_{fix},\ INFO_{CWE})\}$}{$S \gets S \cup \{v_{fix}\}$}
    }
    $V \gets V_{temp}$\\ 
    $i \gets i+1$
}

\Return{$S$}

\end{algorithm}

\subsubsection{\textbf{Method of Code Review}}\label{review}
Code snippets generated or repaired by LLMs need to undergo review to determine whether they are vulnerable to the specified CWEs. For instance, code generated for task \verb|CWE-089/author_1.py| will be examined for the existence of CWE-89 (SQL Injection). \textit{We do not inspect the presence of other CWE vulnerabilities} for (1) it is impossible for any method to encompass the entire spectrum of possible CWEs; (2) according to the design of SecurityEval, the generated code is most likely to exhibit the predefined vulnerabilities; and (3) focusing on predefined vulnerabilities is a common practice in previous studies such as \cite{pearce2022asleep, siddiq2022securityeval, pearce2023examining}. To ensure maximum accuracy, we perform 3 independent rounds of review for each piece of code: 

\textbf{(1) CodeQL} is an open-source semantic code analysis engine developed and maintained by GitHub, capable of detecting code vulnerabilities in various programming languages \cite{codeqldoc}. It is also the tool supporting GitHub Advanced Security features \cite{githubsecurity}. CodeQL officially supports scanning (i.e., provides query scripts) for 29 CWEs and 1 CVE directly related to security in Python \cite{queryscripts}. Of these, 26 CWEs are among the 69 CWEs in SecurityEval, covering 67 out of 121 pieces of its code. We choose CodeQL because of its status as an industry-leading engine for static code analysis and its widespread use in software security research. The version we used is CodeQL CLI v2.16.6. 

\textbf{(2) Bandit} is a semantic code analysis tool specifically designed for Python by PyCQA (an organization for Python code standardization). Bandit is able to traverse code files, build abstract syntax trees, and report potential security vulnerabilities in CWE \cite{banditdoc}. We chose Bandit for our research because it is tailored for Python, aligning closely with the focus of our work. The version utilized in our experiment is Bandit 1.7.8.
% While PyCQA has not officially announced Bandit's coverage for CWE vulnerabilities, our practical experience reveals its capability to identify a range of vulnerabilities that may be overlooked by CodeQL.

\textbf{(3) Manual Review.} Since CodeQL and Bandit do not comprehensively cover all CWEs in the SecurityEval dataset, 4 reviewers were tasked with conducting independent manual code reviews as a supplementary measure. During this process, the reviewers primarily referred to the example insecure code provided by SecurityEval for each task, as well as the CWE definitions available on the official MITRE website\footnote{\href{https://cwe.mitre.org}{https://cwe.mitre.org}}. In cases of disagreement, the final decision was made by an expert in Software Engineering following a group discussion with the reviewers. This approach ensures a thorough evaluation of potential vulnerabilities beyond the scope of automated analysis tools.

We consider a piece of code $x$ to be vulnerable to its corresponding CWE if it has been identified as insecure in any of the three rounds of review, as described in \eqref{eq:standardreview} where $Vul(x)$ denotes whether the code is vulnerable.

\begin{equation}
Vul(x)=\begin{cases}
True\ \ Manual(x)\ ||\ Codeql(x)\ ||\ Bandit(x)\\
False\ \ \textbf{else}
    \end{cases}
    \label{eq:standardreview}
\end{equation}

\subsection{Prompts and Parameters}\label{prompt}
This subsection describes the prompts (in \textit{zero-shot} fashion) and parameters used in our research. \textbf{Due to the large number of cases and the significant effort required for manual review, we only focused on the performance of the LLMs under the following set of prompts}:

\textbf{(1) Prompt for Code Generation.} By referring to the OpenAI official website for guidance \cite{apiintro, promptengineering}, we construct the following prompt for code generation, where \verb|<Code>| denotes a generation task in the SecurityEval dataset:

\textit{``You are a programming assistant, your task is to complete the following Python code according to comments faithfully:} \verb|<Code>|\textit{"}

The above prompt is used in RQ1 and RQ4.

\textbf{(2) Prompt for Vulnerability Detection.}
We refer to relevant work on LLMs for vulnerability detection \cite{fu2023chatgpt, purba23, yin2024multitaskbased} to construct the following prompt used in RQ2, in which \verb|<SN>| and \verb|<Def>| respectively denote the serial number and the definition of the CWE:

\textit{``Your task is to determine whether the following Python code is vulnerable to CWE-}\verb|<SN>|\verb|(<Def>)|\textit{. Return True for vulnerable or False for invulnerable."}

\textbf{(3) Prompt for Vulnerability Repair.}
We refer to the examples for fixing bugs in Python programs on the OpenAI website \cite{pythonbugfixer} to construct the following prompt for code repair which is used in RQ3 and RQ4:

\textit{``You will be provided with a piece of Python code vulnerable to CWE-}\verb|<SN>|\verb|(<Def>)|. \textit{Your task is to generate the complete fixed code."}

\textbf{(4) Parameters.} To strike a balance between the models' creativity and the reproducibility of our work, we did not override the default parameters of the models. For the GPT family, default values of the parameters are \verb|temperature=1|, \verb|top_p=1|, etc. \cite{apiintro}, while for Code Llama they are \verb|temperature=0.1|, \verb|top_p=1|, etc \cite{nvidiacodellama}. Unfortunately, the default values of CodeGeeX2's parameters remain undisclosed. 

We prompt the LLMs to complete all given tasks \textit{in separate conversations.}

\subsection{Experimental Platform}
In this work, we access all LLMs through remote APIs. Results generated by GPT-3.5, GPT-4, and Code Llama are obtained via APIs provided by OpenAI\footnote{\href{https://api.openai.com/v1/chat/completions}{https://api.openai.com/v1/chat/completions}} and NVIDIA\footnote{\href{https://integrate.api.nvidia.com/v1}{https://integrate.api.nvidia.com/v1}}. CodeGeeX2 is accessed through the CodeGeeX extension v2.8.0 for VSCode\footnote{\href{https://marketplace.visualstudio.com/items?itemName=aminer.codegeex}{https://marketplace.visualstudio.com/items?itemName=aminer.codegeex}}. CodeQL, Bandit, and all the Python scripts used in our work are run locally on a single desktop-class PC equipped with an Intel i5-11300H processor and 16 GB DDR4 RAM.

% \section{RQ1: How do LLMs perform when generating Python code in security-sensitive scenarios?}
\section{STUDY OF RQ1}

\subsection{Overview}

In this section, we have GPT-3.5, GPT-4, Code Llama, and CodeGeeX2 generate code for tasks in the SecurityEval dataset with the prompt and parameters described in \ref{prompt}. Overall, 484 pieces of code have been generated (121 pieces by each of the 4 LLMs). We then automatically and manually review all the generated code with the method described in \ref{review}. 

\subsection{Results and Analysis}

The detailed visual representation of the performance of the 4 LLMs on SecurityEval can be accessed in our repository. As a summary, Table \ref{tab:Q1_result} shows the number and percentage of insecure code pieces generated by each model.

\begin{table}[t]
\caption{Vulnerable code pieces generated by the LLMs}
\centering
\begin{tabular}{c|l|c|l}
\hline
Model&\#Vul&Model&\#Vul\\
% \textbf{GPT-3.5}&\textbf{GPT-4}&\textbf{Code Llama}&\textbf{CodeGeeX2}&\textbf{Avg.} \\
\hline\hline
GPT-3.5&92/121 (76.0\%)&Code Llama&91/121 (75.2\%)\\
\hline
GPT-4&91/121 (75.2\%)&CodeGeeX2&95/121 (78.5\%)\\
\hline
\cellcolor{light-gray}\textbf{Avg.}&\cellcolor{light-gray}369/484 (76.2\%)&&\\
% \multicolumn{2}{c|}{\cellcolor{light-gray}}
\hline
\end{tabular}
\label{tab:Q1_result}
% \vspace{-0.4cm}
\end{table}

% \subsection{Result Analysis}

Despite their advanced capability in generating functionally correct Python code \cite{humaneval} \cite{zheng2023codegeex}, all four models performed poorly in generating secure code on SecurityEval, with an average of 76.2\% of the generated code being insecure. This highlights a significant challenge in using current LLMs for secure code generation and underscores the need for prompting them to detect and fix the vulnerability in their self-generated code, as studied in RQ2, RQ3, and RQ4.

As the emphasis of our work is not on the security of LLM-generated code, which has been extensively addressed in prior research, \textbf{a detailed analysis of the performance of the 4 LLMs on SecurityEval is provided in our repository for readers who are interested}.

\begin{table*}[ht]
\caption{GPT-3.5's and GPT-4's Accuracy in identifying vulnerabilities in code generated by the 4 LLMs}
\centering
\begin{tabular}{c|l|l|l|l|c}
\hline
\multirow{2}{*}{Model for Detection} & \multicolumn{4}{c|}{Model for Generation} & \multirow{2}{*}{\textbf{Avg.}}\\ 
\cline{2-5}
\multicolumn{1}{c|}{} & GPT-3.5 & GPT-4 & Code Llama & CodeGeeX2 & \\
\hline\hline
GPT-3.5&58/121\ (47.9\%) & 52/121\ (43.0\%) & 52/121\ (43.0\%) & 49/121\ (40.5\%) &211/484 (43.6\%)\\
\hline
GPT-4 &91/121\ (75.2\%) & 90/121\ (74.4\%) & 87/121\ (71.9\%) & 93/121\ (76.9\%) &361/484 (74.6\%)\\
\hline
\end{tabular}
\label{tab:agree_rate}
% \vspace{-0.3cm}
\end{table*}

% \section{RQ2: How effective are LLMs in identifying LLM-generated code vulnerabilities?}
\section{STUDY OF RQ2}

\subsection{Overview}

In this section, we investigate whether large language models are qualified code self-reviewers by prompting them to identify vulnerabilities in code produced by themselves. In specific, we ask them whether the specified CWE weakness exists in the code they have generated, and compare their judgments with \textit{the review results gained in RQ1 which serve as the ground truth}. Detailed prompts are presented in \ref{prompt}.

Instead of using all four models, we use only GPT-3.5 and GPT-4, as they have demonstrated the ability to generate coherent responses to our queries for vulnerability detection. Code Llama and CodeGeeX2 are excluded because, during our preliminary tests for the capability to detect, they often generate nonsensical responses, regardless of our command to identify weaknesses (part of their responses can be found in our public repository for reference). Overall, GPT-3.5 and GPT-4 generate 968 pieces of judgment (484 by each of the 2 models). 

\subsection{Results and Analysis}
We assessed the correctness of a LLM's judgment by checking whether it is in agreement with the established ground truth. Equation \eqref{eq:acc} illustrates the method in a formalized way, in which $Tech$ represents either GPT-3.5 or GPT-4. 

\begin{equation}
Acc_{Tech}(x)=\begin{cases}
True& Tech(x) \ == \ Vul(x)\\
False& \textbf{else}
    \end{cases}
\label{eq:acc}
\end{equation}

Detailed results of this section are available in our repository, while Table \ref{tab:agree_rate} provides a statistical summary of the accuracy of GPT-3.5 and GPT-4 in detecting vulnerabilities in code produced by all four LLMs.

% \subsection{Result Analysis}

On average, GPT-3.5 achieves 43.6\% accuracy, which is comparable to random-guess baseline ($\sim$50\%). This suggests that GPT-3.5's ability to detect vulnerabilities is of limited practical value. GPT-4, on the other hand, achieves an average accuracy of 74.6\%, demonstrating its superior ability to understand and analyze code. The experimental results are largely consistent with previous related work based on real-world code \cite{purba23}.

In addition to accuracy, false positive rate (FPR) is another crucial metric for assessing the reliability of a detection technique. By \textit{only considering the results of the manual review in RQ1 as the ground truth}, we determine a judgment as false positive using \eqref{eq:false-positive}, in which $FP_{Tech}(x)$ denotes whether a result of $Tech$ (GPT-3.5, GPT-4, CodeQL, or Bandit) is false positive. 

\begin{equation}
FP_{Tech}(x)=\begin{cases}
True& Tech(x) \ \&\&\ !Manual(x)\\
False& \textbf{else}
    \end{cases}
\label{eq:false-positive}
\end{equation}

\begin{figure}[t]
    \centering
    \includegraphics[width=0.8\linewidth]{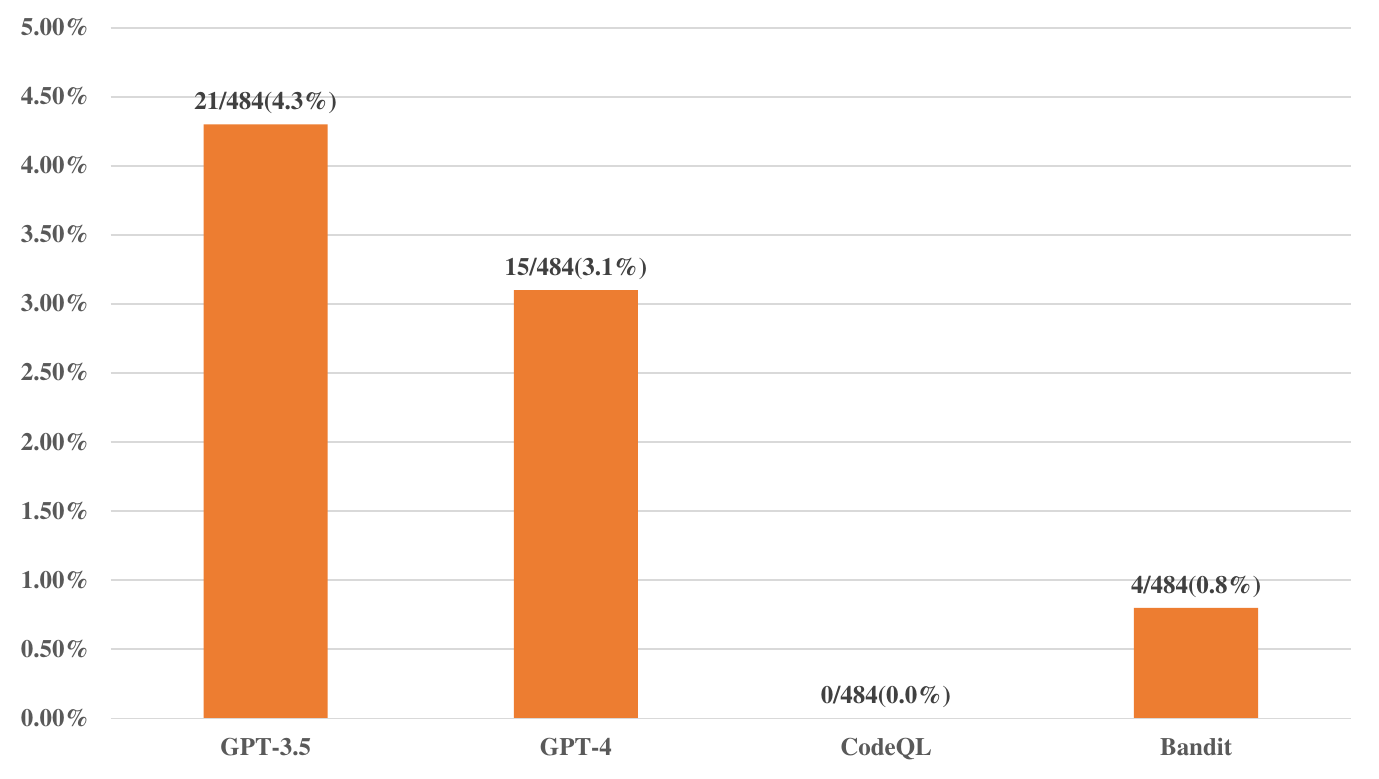}
    \caption{False positive rates of the 4 studied techniques}
    \label{fig:FPR}
    % \vspace{-0.3cm}
\end{figure}

Fig. \ref{fig:FPR} presents the false positive rates of GPT-3.5, GPT-4, CodeQL, and Bandit. The FPRs of CodeQL and Bandit serve as baselines for assessing the trustworthiness of GPT-3.5 and GPT-4 as code reviewers. Overall, the FPR of GPT-3.5 reaches 4.3\%, while that of GPT-4 is 3.1\%, both of which are unacceptably high. In contrast, CodeQL and Bandit yield relatively low false positive rates (below 1\%). This indicates that both tested LLMs have a significant likelihood of incorrectly identifying secure code as vulnerable. Consequently, \textbf{neither GPT-3.5 nor GPT-4 can be relied upon for accurate vulnerability detection in the code they generated.}  

% \section{RQ3: How Effective are LLMs in Repairing LLM-generated Code Vulnerabilities?}

\begin{table*}
\caption{GPT-3.5's and GPT-4's success rates of repairing code generated by the 4 LLMs in a single attempt ($R_{fix}^{1}$)}
\centering
\begin{tabular}{c|l|l|l|l|c}
\hline
\multirow{2}{*}{Model for Repair} & \multicolumn{4}{c|}{Model for Generation} & \multirow{2}{*}{\textbf{Avg.}}\\ 
\cline{2-5}
\multicolumn{1}{c|}{} & GPT-3.5 & GPT-4 & Code Llama & CodeGeeX2 & \\
\hline\hline
GPT-3.5 & \textbf{28/92 (30.4\%)} & 30/91 (33.0\%) & 31/91 (34.1\%) & 34/95 (35.8\%) &123/369 (33.2\%)\\
\hline
GPT-4 & 56/92 (60.9\%) & \textbf{50/91 (55.0\%)} & 58/91 (63.7\%) & 56/95 (58.9\%) & 220/369 (59.6\%)\\
\hline
\end{tabular}
\label{tab:fix_result}
% \vspace{-0.3cm}
\end{table*}

\section{STUDY OF RQ3}

\subsection{Overview}

In this section, we investigate whether large language models are capable of effective code self-repair by prompting them to fix weaknesses in previously identified vulnerable code snippets. Subsequently, we manually and automatically examine whether the pieces are fixed with the method described in \ref{review}. 

Only GPT-3.5 and GPT-4 are included in this analysis, for the same reason in RQ2. Prompts used in this part have been detailed in \ref{prompt}. In total, GPT-3.5 and GPT-4 generated 738 pieces of repaired code (369 pieces of vulnerable code generated by the 4 LLMs in RQ1, repaired respectively by GPT-3.5 and GPT-4), all of which underwent both manual and automated review as outlined in \ref{review}.

\subsection{Results and Analysis}
Detailed results of this RQ can be accessed in our repository, which includes visualizations of GPT-3.5's and GPT-4's performance in repairing vulnerable code with each CWE.

Table \ref{tab:fix_result} presents the success rates of repair computed using \eqref{eq:repair_1}, in which $N_{vul}$ denotes the number of vulnerable code snippets and $N_{fix}^{1}$ denotes the number of those who were successfully repaired (superscript $1$ stands for one single attempt). 
\begin{equation}
\text{$R_{fix}^{1}$} = \frac{\text{$N_{fix}^{1}$}}{\text{$N_{vul}$}}\times 100\%
\label{eq:repair_1}
\end{equation}

% \subsection{Result Analysis}

\subsubsection{Statistical Analysis}

Table \ref{tab:fix_result} demonstrates that GPT-3.5 and GPT-4 are capable of repairing a range of LLM-generated insecure code when provided with a description of the CWE type. Notably, GPT-4 performs significantly better than GPT-3.5, with nearly twice the success rate of repair. Although there are no pre-existing techniques as baselines to compare with (to the best of our knowledge, there has been no automated technique such as APR tools that can effectively fix security vulnerabilities in Python programs), \textbf{it can be concluded that advanced large language models such as GPT-4 have a promising level of ability to repair vulnerabilities in the code generated by themselves or other LLMs.}

As emphasized in Table \ref{tab:fix_result}, GPT-3.5 achieves a success rate of 30.4\% when attempting to fix its own generated code, marking its poorest performance across all code repair tasks. Similarly, GPT-4 has its lowest success rate of 54.3\%  when fixing the code it generated. \textbf{This evidence suggests that large language models tend to experience a decline in performance when attempting to fix vulnerabilities generated by themselves. However, this finding needs to be tested across a broader spectrum of scenarios to ensure its validity.} This insight is particularly interesting and noteworthy, as it sheds light on the limitations of LLMs in improving the content generated by themselves. 

\subsubsection{Scenario-relevant Analysis}

\begin{figure}[t]
    % \vspace{-0.2cm}
    \centering
    \includegraphics[width=0.58\linewidth]{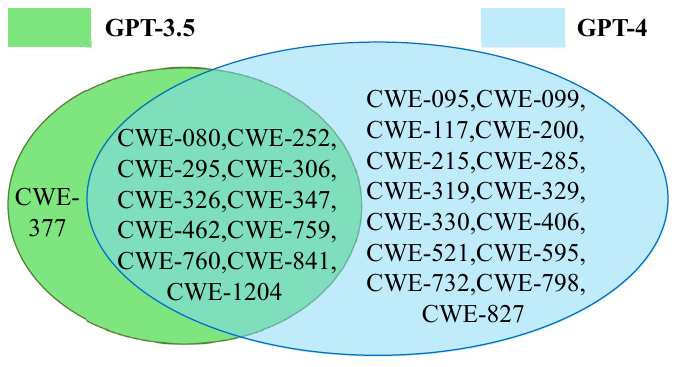}
    \caption{The Venn diagram of CWEs where all vulnerable code generated by the 4 LLMs is fixed by GPT-3.5 or GPT-4}
    \label{fig:good_result2}
\end{figure}

Fig. \ref{fig:good_result2} illustrates the CWE scenarios in which GPT-3.5 and GPT-4 successfully repaired all vulnerabilities in the code generated by the four LLMs. GPT-3.5 successfully repaired all the vulnerable code snippets in 12 CWE categories, while GPT-4 achieved this in 26 CWE categories, with 11 CWEs being common to both models. It is evident that GPT-3.5's coverage of success repair is nearly a subset of GPT-4's, indicating that GPT-4 was able to address significantly more CWE scenarios than GPT-3.5. 
% This highlights GPT-4's superior capability in repairing security vulnerabilities compared to GPT-3.5, reaffirming its effectiveness in enhancing code security. 

% \section{RQ4: How Effective Is an Iterative Strategy in Improving LLMs' Repair Capability?}
\section{STUDY OF RQ4}

\subsection{Overview}

\begin{figure*}[t]
    % \vspace{-0.2cm}
    \centering
    \includegraphics[width=\linewidth, height=4.4cm]{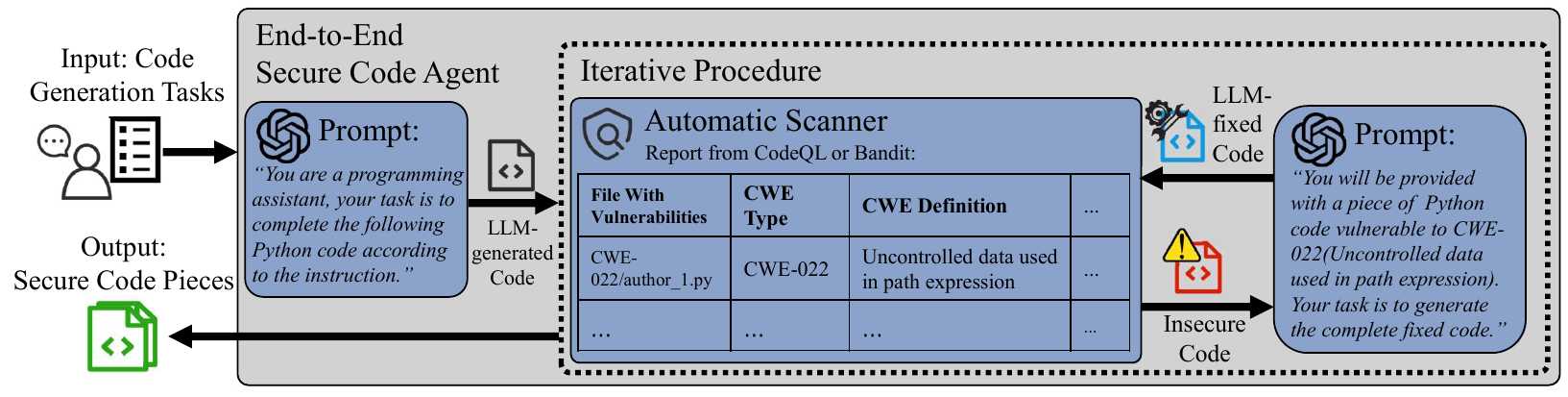}
    \caption{The design of our end-to-end secure code agent}
    \label{fig:tool}
\end{figure*}

RQ2 and RQ3 respectively represent the stages of code review and refactoring of the end-to-end procedure. However, one single round may not be sufficient to address all security issues, as indicated in RQ3. Therefore, in RQ4, we investigate the effectiveness of implementing an iterative strategy that repeatedly conducts vulnerability detection and repair to enhance LLMs' repair capability. With multiple rounds of attempts and specific feedback provided, we expect LLMs to produce source code with less vulnerabilities.

Since this process produces a large amount of generated and repaired code, we developed an automated tool implementing the algorithm described in Algorithm \ref{algorithm}. As depicted in Table \ref{fig:tool}, our tool for RQ4 consists of a code generator, a vulnerability scanner, and a vulnerability repairer. The roles of the code generator and repairer are performed by the LLMs being evaluated. \textit{Instead of using LLMs as scanners, we utilize reliable semantic code analysis engines (CodeQL and Bandit) in the tool, given that GPT-3.5 and GPT-4 have demonstrated their inability to correctly identify vulnerabilities in RQ2.} We only manually examine the final pieces of code generated after all iterations.

The tool takes as input files containing code generation tasks. It first calls the API of the LLM which serves as the generator to produce code snippets based on the task description. Its scanner then scans the generated code files for weaknesses according to CWE specifications. A piece of generated code is deemed vulnerable by the tool if reported as insecure by either of the two engines.
% % \vspace{-0.05cm}
% \begin{equation}
% Vul(x)=\begin{cases}
% True& Codeql(x)\ ||\ Bandit(x) \\
% False& \textbf{else}
%     \end{cases}
% \end{equation}
% % \vspace{-0.2cm}

Code snippets free of weaknesses will be output as secure code, while those found to have vulnerabilities are returned to the LLM who serves as repairer for remediation, given the CWE information of the weaknesses (provided by CodeQL or Bandit). This scan-and-repair process is conducted in an iterative manner until all code is regarded as secure code or the predefined maximum iterations are reached. This tool is available in the public repository of our work.

As outlined in \ref{review}, CodeQL is capable of scanning for 26 CWEs in SecurityEval, corresponding to 67 code generation tasks, while Bandit's coverage remains undisclosed to us. Therefore, \textit{to align with the detection capabilities of the automated analysis tools, we used only the 67 out of 121 code generation tasks from the SecurityEval dataset that are directly analyzable by CodeQL in RQ4}. This approach ensures that predefined CWE risks can be detected by at least one of the two analysis tools.

Four experimental setups were designed for RQ4: GPT-3.5 self-repairing code from GPT-3.5, GPT-4 self-repairing code from GPT-4, GPT-3.5 cross-repairing code from GPT-4, and GPT-4 cross-repairing code from GPT-3.5. We conducted 5 independent experiments for each setup to mitigate random factors. In total, the 20 experiments produced and automatically examined about 3,500 code pieces. Finally, we manually reviewed the code generated from the last repair round.

\subsection{Results and Analysis}
\begin{figure}[t]
    \centering
    % \vspace{-0.1cm}
    % \setlength{\abovecaptionskip}{-0.03cm}
    \includegraphics[width=\linewidth, height=6cm]{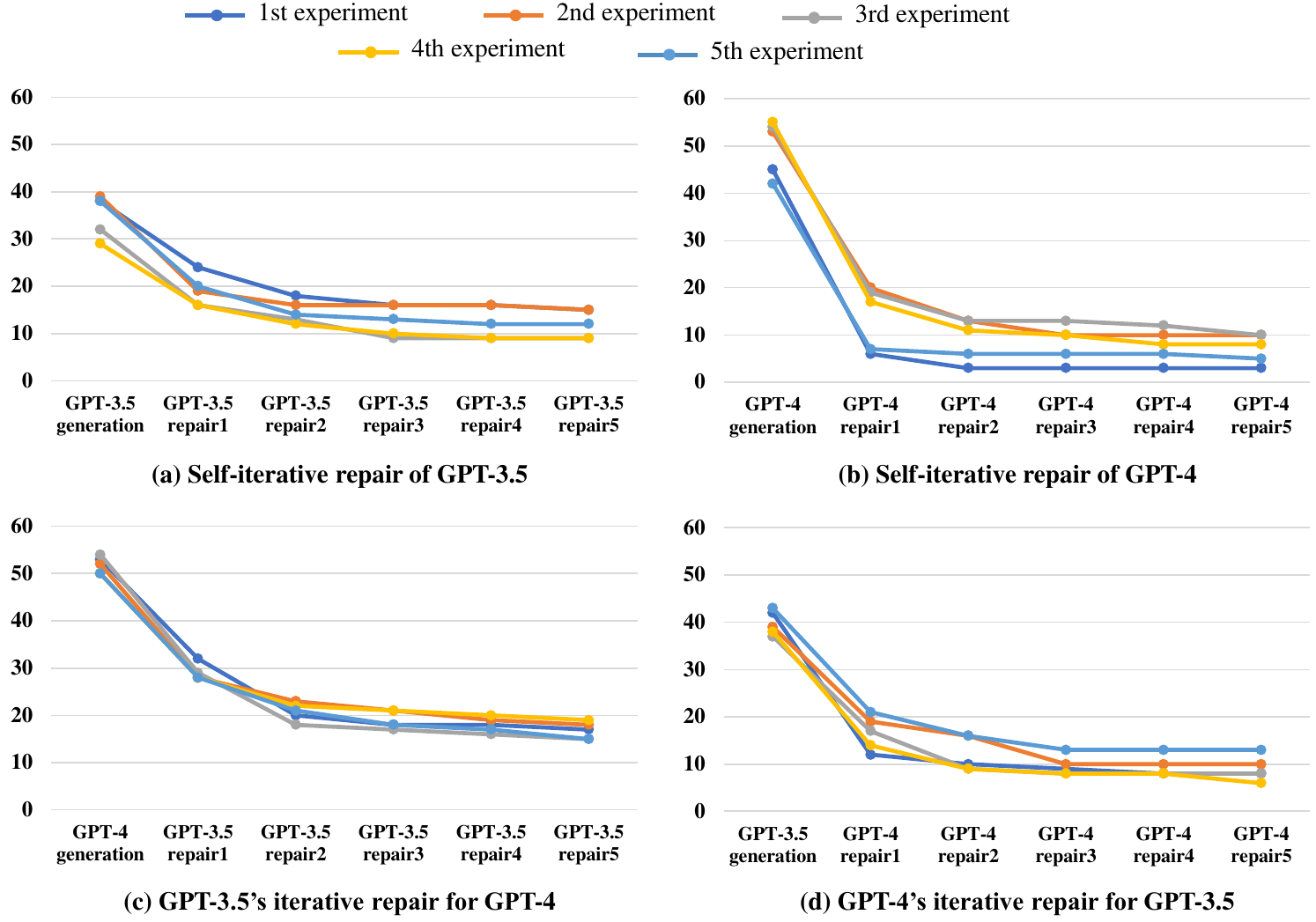}
    \caption{Line chart depicting the number of vulnerable code pieces identified by CodeQL or Bandit across each iteration}
    \label{fig:RQ4_result}
    % \vspace{-0.5cm}
\end{figure}

Fig. \ref{fig:RQ4_result} depicts the results of repair across each iteration using the tool we developed. 

As a quantitative summary of Fig. \ref{fig:RQ4_result}, Table \ref{tab:iterate_result} presents the averaged cumulative success rates after five consecutive repair iterations. These rates are computed using \eqref{eq:repair_5}, where $\overline{N}_{vul}$ represents the mean of the number of vulnerable code snippets initially identified by CodeQL or Bandit following the code generation phase of the 5 independent experiments of each setup, and $\overline{N}_{fix}^{5}$ denotes the mean of number of those who were successfully repaired within or by the conclusion of the 5th iteration (superscript 5 stands for five iterative attempts).
\begin{equation}
\text{$\overline{R}_{fix}^{5}$} = \frac{\text{$\overline{N}_{fix}^{5}$}}{\text{$\overline{N}_{vul}$}}\times 100\%
\label{eq:repair_5}
\end{equation}

% \vspace{-0.5cm}
\begin{table}[t]
\renewcommand{\arraystretch}{0.9}
\caption{Averaged cumulative success rates ($\overline{R}_{fix}^{5}$)}
\centering
\begin{tabular}{c|l|l}
\hline
\multirow{2}{*}{Model for Repair}
& \multicolumn{2}{c}{Model for Generation}\\
\cline{2-3}
\multicolumn{1}{c|}{} & GPT-3.5 & GPT-4  \\
% \morecmidrules\cmidrule(lr){2-3} & GPT-3.5 & GPT-4  \\
\hline\hline
GPT-3.5 & 65.9\% & 67.6\% \\
\hline
GPT-4 & 77.4\% & 85.5\% \\
\hline
\end{tabular}
\label{tab:iterate_result}
\end{table}
% \vspace{-0.4cm}
% 176,60;249,36;259,84;199,45

% \subsection{Result Analysis}

Across the 20 experiments (5 experiments for each of the 4 setups), CodeQL and Bandit initially identified an average of about 45 vulnerable code pieces. As the iterative repair process progressed, the number of detected vulnerabilities significantly decreased. After the final iteration, only an average of 10 code pieces are still found to have weaknesses. On average, GPT-3.5 and GPT-4 successfully repaired 65.9\% $\sim$ 67.6\% and 77.4\% $\sim$ 85.5\% of initial vulnerable code pieces, respectively. These success rates are significantly higher than those observed in RQ2, where only a single repair attempt was made. \textbf{The results highlight feedback-driven self-iterative repair as a promising approach for LLMs to enhance security in the code they have generated.}

Observed from Fig \ref{fig:RQ4_result}, the reduction slows down considerably after the second repair iteration, indicating that the efficacy of iterative repairs is approaching its limit. While iterative repair does improve the success rate of repair, it becomes evident that \textbf{excessive iterations contribute little to enhancing the overall repair efficiency}. Additionally, excessive iterations can be time-consuming and expensive, which highlights a critical trade-off between code functionality, security, and development efficiency. 

\section{IMPLICATIONS AND DISCUSSIONS}
% Our study identifies several important implications and suggestions for the research of large language models for code and vulnerability repair.

% \subsection{The need for a larger coverage of semantic code analysis engines for vulnerability detection}

% Semantic code analysis engines such as CodeQL are renowned for their reliability in identifying vulnerabilities. These tools, driven by manually written query scripts targeting specific weaknesses, typically exhibit low false positive rates in practical scenarios. Therefore, they are widely used in research for software security \cite{pearce2022asleep, pearce2023examining}. However, current engines fall short in terms of their coverage (i.e., the number of detectable CWEs). Consequently, many studies, including ours, resort to manual code review to ensure comprehensive coverage, albeit at the expense of time and effort. Expanding the coverage of these analysis engines would significantly boost the efficiency and reproducibility of relevant research endeavors.

\subsection{LLMs' awareness of security risks}

In RQ1, it was observed that LLMs produced a significant amount of insecure code when tasked with scenarios involving specific security risks. However, the result does not necessarily imply that LLMs are incapable of generating more secure code, for they were able to correct many of the vulnerabilities present in their generated code when prompted to do so. We posit that the production of vulnerable code by LLMs largely stems from their lack of awareness regarding security issues, as they primarily prioritize fulfilling functional requirements. In real-world scenarios, software developers do not always provide LLMs with information about relevant risks. Therefore, it is crucial to enhance the scenario-relevant security awareness of LLMs, particularly that of code language models which are designed for code-related tasks. Additionally, it is recommended that users explicitly include brief descriptions of potential security weaknesses in prompts to guide LLMs in preventing them.

\subsection{Self-repair ``blind spots" of LLMs}

An intriguing observation from RQ3 is that both GPT-3.5 and GPT-4 achieve their lowest success rates when repairing code generated by themselves (as revealed in Table \ref{tab:fix_result}). This suggests that similar to human programmers who tend to overlook the weaknesses in their self-written source code, LLMs also exhibit ``blind spots" in code self-repairing. We presume that the phenomenon exists because LLMs are too dependent on the programming patterns learned from their training stage that they tend to ``insist" on these patterns rather than exploring alternative approaches, leaving vulnerabilities unfixed when prompted to address self-produced weak code. Conversely, when fixing insecure code generated by other models, a large language model can leverage its unique patterns to address weaknesses that other LLMs may overlook, resulting in a slightly higher success rate of repair.

% \subsection{General large language models versus code language models}

% In the studies of RQ2, RQ3, and RQ4, we excluded Code Llama and CodeGeeX2 due to their inability to generate responses coherent with our prompts. Although these two language models achieve remarkable results on code generation tasks \cite{humaneval}, they perform poorly on other code-related tasks such as vulnerability detection and repair, often generating either garbled code or self-conflicting responses. In contrast, general-purpose large language models like GPT-3.5 and GPT-4 can comprehend prompts for detection and repair, thus generating satisfactory results. The disparity may be attributed to the fact that larger-scale language models are trained on extensive natural language datasets, which enables them to comprehend prompts and generate coherent responses. Accordingly, future frameworks for automated secure code construction may either leverage the advantages of general-purpose large language models or utilize specialized models that have been fine-tuned for vulnerability detection and repair.

\section{THREATS TO VALIDITY}
 
% \textit{Reproducible Code Generation:}
% As generative models, LLMs used in this work are unable to produce completely reproducible output. Given the time-consuming nature of manual code review, we instructed the LLMs to generate only one output for each task (otherwise the amount of code to review would multiply). Consequently, our results may be affected by random factors, as LLMs can produce different outputs when given the same prompt. We contend that such influence is minimal, as all results were generated under default parameters with medium model temperatures. To alleviate doubts, we particularly had GPT-4 generate three parallel outputs for each generation task in RQ1. Manual review confirms that GPT-4 consistently produced similar results across these outputs\footnote{The code and review results of this process are included in our public repository}. 

\textit{Choice of the Dataset:}
The SecurityEval dataset used in our work was released two years prior to our research and might have been included in the training data of LLMs. Despite this possibility, all 4 tested LLMs exhibit poor performance in terms of security quality when assessed against this benchmark. Therefore, the security problem of large language models for code remains an open challenge and our conclusions drawn from the experiments with SecurityEval maintain their validity and relevance. 

\textit{Limitations of Code Review:}
While manual code review is a prevalent practice in the software engineering community, it does not guarantee 100\% accuracy because of subjective factors. Moreover, the security of code remains an open question, as standards for secure coding may evolve with the emergence of new attack methods or updates to library functions (e.g., Python standard modules). Nevertheless, we are confident that our method for code review described in \ref{review} provides largely reliable judgments. Additionally, we make the review results publicly accessible in our repository.

\textit{Limitations of Experimental Design:}
Due to time constraints, we did not fully explore the potential of LLMs in terms of prompt engineering. It is conceivable that prompts with more detailed task descriptions, such as specifying the row and column numbers of vulnerabilities or using different prompting styles, could influence the performance of LLMs. 

\section{RELATED WORK}

\textbf{(1) Vulnerabilities in LLM-generated code.}\ 
% With large amounts of code being generated by LLMs and deployed into production environments every day, their security has become a significant concern for both academia and industry. 
Pearce et al. were among the first to evaluated the security of C and Python code generated by GitHub Copilot across 25 CWE cases, finding that 40\% of the code was vulnerable \cite{pearce2022asleep}. Furthermore, \cite{khoury2023secure, asare2023github, sandoval2023lost, fu2023security, liu2024no} also examine LLMs' ablity to generate secure code in various settings. A more recent study by Tihanyi et al. examined the security of C code generated by GEMINI-pro, GPT-4, and other models, revealing that at least 63.47\% of the generated programs were vulnerable \cite{tihanyi2024neutral}, which is a number close to our findings on Python. These results highlight the inability of current LLMs to consistently generate secure code without elaborately designed prompts.

\textbf{(2) LLMs for detecting vulnerabilities in real-world code.}\ 
Recent research has increasingly focused on the direct application of LLMs in enhancing code security, particularly in the areas of vulnerability detection and repair \cite{zhou2024large}. Fu et al. investigated the ability of LLMs to detect and classify weaknesses in real-world code \cite{fu2023chatgpt}. Purba et al. applied 4 well-known LLMs to detect vulnerabilities in 2 datasets. They found a significant performance gap between the studied LLMs and static analysis tools, primarily due to LLMs' high false positive rates \cite{purba23}, a result consistent with our conclusion in RQ2. Other research also highlighted the limitations of current LLMs in vulnerability detection compared to static analysis tools or specially trained, deep learning-based models \cite{steenhoek2024comprehensive, yin2024multitaskbased}. Contrary to these findings, some researchers have observed LLMs' superiority in specific experimental settings \cite{zhou2024largedetect}\cite{akuthota23}. Ullah et al. designed SecLLMHolmes, an automated evaluation framework that investigates LLMs' reliability in identifying and reasoning about security-related bugs \cite{ullah2024llms}. Yang et al. proposed a novel framework that reinforces LLMs with deep learning-based techniques, achieving state-of-the-art performance in vulnerability detection \cite{yang2024dlap}. 

\textbf{(3) LLMs for repairing vulnerabilities in real-world code.}\ 
Research results vary regarding the efficacy of large language models in vulnerability repair. A study by Pearce et al. indicates that LLMs are promising zero-shot vulnerability fixers \cite{pearce2023examining}. Wu et al. highlighted the advantages of Codex over traditional deep learning-based repair tools for addressing CWE weaknesses in Java code \cite{Wu_2023}. Le et al. concluded that ChatGPT provides satisfactory repair results for JavaScript code when detailed descriptions of the vulnerabilities are given \cite{Le_2024}. Ahmad et al. found that LLMs such as Codex can effectively fix security bugs in Verilog, a hardware programming language \cite{Ahmadverilog}. Conversely, Fu et al. reported that LLMs such as GPT-3.5 underperform compared to traditional models like CodeBERT in vulnerability repair tasks \cite{fu2023chatgpt}. 

It is noteworthy that most previous research focuses solely on LLMs' efficacy in detecting and fixing vulnerabilities in real-world, manually written code. \textbf{While these studies provide valuable insights, they do not fully reveal the potential of LLMs to be end-to-end secure code agents who must detect and repair vulnerabilities in the code snippets generated by themselves. This underscores the need for our research specifically targeting the security of \textit{LLM-generated} code.}

\section{CONCLUSIONS AND PERSPECTIVES}
In this paper, we seek an answer to the question of how well LLMs serve as end-to-end secure code agents for Python. We first investigate the vulnerabilities in Python source code generated by GPT-3.5, GPT-4, Code Llama, and CodeGeeX2 on the SecurityEval benchmark. Subsequently, we explore LLMs' potential to independently enhance the security of the code through code self-review and vulnerable code self-repair. Overall, we manually or automatically examined around 4,900 code pieces. 

Our study reveals several key findings: (1)\ LLMs tend to generate insecure code in security-critical programming tasks because of shortage of scenario-relevant awareness of potential risks; (2)\ LLMs such as GPT-3.5 and GPT-4 are not capable of accurately identifying vulnerabilities in the source code they produce, primarily due to their high false positive rates; (3) advance LLMs can achieve up to a 60\% success rate repairing insecure code generated by other LLMs, but they exhibit relatively poor performance when repairing self-produced code; (4)\ Leveraging semantic code analysis engines, a feedback-driven self-iterative repair approach significantly enhances the security of LLM-generated code.

\textbf{While we hold the belief that future LLMs have the potential to produce secure code in an end-to-end fashion, current models are unable to serve as reliable secure code agents without assistance from established tools like semantic code analysis engines.}  

Our study also leads us to the following viewpoints:

(1) We recommend that software developers explicitly highlight potential security risks when instructing large language models to generate source code.

(2) We suggest augmenting LLMs for code with scenario-specific fine-tuning to enhance their security awareness to mitigate potential vulnerabilities. 

(3) We advocate for efforts to expand the coverage of semantic code analysis engines such as CodeQL, increasing their capability to detect a wider range of CWEs for the benefit of automated and reproducible research.

\section{ACKNOWLEDGMENTS}
In the preparation of this manuscript, ChatGPT was utilized to refine certain sentences to enhance their clarity, precision, and academic rigor. However, the core ideas, analysis, and conclusions remain entirely the authors' own. The use of AI-assisted editing was strictly limited to linguistic improvements and did not influence the scholarly integrity or originality of the research.

\newpage
\bibliographystyle{ACM-Reference-Format}
\bibliography{reference}

\end{document}